\documentclass{PoS}
\usepackage{listings}
\usepackage[T1]{fontenc}
\usepackage[utf8]{inputenc}
\usepackage{dirtree}
\usepackage{caption}
\usepackage{subfig}

\title{A Prototype for the Cherenkov Telescope Array Pipelines Framework: Modular Efficiency Simple System (MESS)}

\ShortTitle{MESS}

\author{
    \speaker{Ramin Marx}\\
    Max-Planck-Institut für Kernphysik, PO Box 103980, 69029 Heidelberg, Germany\\
    E-mail: \email{ramin.marx@mpi-hd.mpg.de}
}

\author{
    Raquel de los Reyes\\
    Max-Planck-Institut für Kernphysik, PO Box 103980, 69029 Heidelberg, Germany\\
    E-mail: \email{raquel.de.los.reyes@mpi-hd.mpg.de}
}

\abstract{
    The Cherenkov Telescope Array (CTA) is a ground-based $\gamma$-ray observatory that will observe the full sky
    in the energy range from 20 GeV to 100 TeV from facilities in both hemispheres.
    It is proposed to consist of more than 100 telescopes, producing large amounts of data.
    Apart from the storage system, there are also requirements on the software framework to allow efficient data processing, i.e. robustness, execution speed and coding efficiency.
    This contribution will present a plain and simple pipeline framework design prototype for CTA that builds upon well-known tools,
    allowing the users to focus on physics problems without learning complicated software paradigms.
}

\FullConference{
    The 34th International Cosmic Ray Conference,\\
    30 July- 6 August, 2015\\
    The Hague, The Netherlands
}

\begin{document}

\section{Introduction}

CTA will be the first open observatory of very-high-energy $\gamma$-rays.
It will be the successor of the current generation of ground-based imaging atmospheric Cherenkov telescope (IACT) experiments.
Arrays may combine up to six types of telescope and up to seven types of cameras.
This design exceeds the dimension and complexity of the current IACT experiments, which are formed by a maximum of 5 telescopes and with no more than two telescope and camera types.
The telescopes will record Cherenkov light coming from the extensive air showers (EAS) produced by primary $\gamma$-rays and, mostly, by cosmic rays (CR).
The high expected trigger rates of several tens of kHz, together with the $\sim 10^3$ to $10^4$ pixels per camera, will lead to huge CTA raw data rates~\cite{MC}.
These data rates must be processed by the CTA Pipelines~\cite{DM} during the expected life time ($\sim$ 30 years) of the CTA Observatory (CTAO).
This contribution presents the Modular Efficient Simple System (MESS), a CTA Pipeline prototype.
It takes into account the data challenges of the CTAO:
\begin{itemize}
    \item The open observatory condition of CTA requires a more robust framework compared to current experiments.
        CTAO has to provide consistent and reproducible results to the astronomical community.
    \item Due to the long lifetime of the observatory, long-term supported libraries/system will minimise the software maintenance costs.
    \item The diversity of technology within the CTAO will demand a clear modularity between the different software components.
        Being able to run a variety of pipelines without recompiling any program is an advantage and increases the reproducibility of results.
\end{itemize}
The next sections will describe the MESS prototype and preliminary results when applied to CTA Monte Carlo (MC) data.

\section{Software framework}
The idea behind this proposal is that a software framework should allow developers to concentrate on their algorithms and not care about learning new paradigms and loose time doing redundant work.
The MESS software framework provides a robust library with clean interfaces, minimal dependencies and complexity, built with proven and well-known systems.
It comes with tools for automating the redundant development processes and for creating documentation.

\subsection{Dependencies} Dependencies should be minimised, because the fewer there are, the easier it is to build and maintain the software.
The MESS library is written in plain C and has no dependencies except \textsc{cfitsio}~\cite{cfitsio}, which is used for storing tabular data in FITS~\cite{FITS} files.
Whenever required, different libraries and programs can be involved, but such dependencies are only per single module (shared object file) and do not affect the rest of the system.
The module \textit{readctamc}, which reads raw CTA MC data,
needs to be linked to \texttt{hessioxxx} (a library that is part of the \textsc{simtelarray}~\cite{simtelarray} package),
which also does the pixel calibration.
Other MESS programs need to be linked to \textsc{gsl}~\cite{gsl} and \textsc{plplot}~\cite{plplot} to allow plotting histograms and displaying events.

\subsection{Complexity} Library functions should be orthogonal to each other and it should be possible to combine them in a coherent and straightforward way.
Few different data types should be enough to represent the problems.
The current MESS framework version, with basic IACT analysis algorithms implemented, has 4000 lines of code (without comments), which makes it easy to understand and to maintain.
The library and basic modules are compiled in less than 3 s.

\subsection{Build System} MESS uses a global Makefile that dynamically includes all Makefiles in the subdirectory, in order to keep dependencies separated.
When a new module or a new program is added, it is enough to put just its name into the appropriate Makefile.
This way, external libraries and readers/writers for external formats can be integrated in a clean way.

\subsection{Autogenerated header files and documentation}
Writing and maintaining header files is not necessary anymore:
MESS provides the program \texttt{c2h}, which scans through all source code files (\texttt{.c/.cxx}) and creates header files (\texttt{.h}) from them,
including comments above functions, variables, type definitions etc.
Another program in MESS, \texttt{h2txt}, reads these autogenerated header files and converts them to text files in markdown format.
Markdown is like text, but has a few (human readable) tags to allow conversion to nicer looking HTML with optional bold/italic script, lists, images, links and inline code.
The MESS program \texttt{txt2html} does this conversion, adding support for code blocks and Latex formulas on top of markdown.
So the developer only cares about the \texttt{.c/.cxx} files and all redundant work is automatically done.

\subsection{Versioning} Whenever the MESS library is built, \texttt{git}, which is used as version control system for the MESS code, is queried to return the current version of the commit from which the library is to be built.
That data along with time and date are then written into an automatically created file, which is linked into the MESS library.
All programs linked to the MESS library can now query these data, so the developer does not need to keep track of the versions of his software, and in bug reports, the users can provide the version of the library.

\subsection{Logging} MESS provides an infrastructure for global logging and per-module logging, so each module can have its own log file and log level, and it can also write to the global logger.

\section{Data Structures}
MESS uses simple C structures, so developers can write clean code instead of using nested getter/setter methods of classes.
Porting code, for example to GPUs, and writing wrappers for other languages is also much easier.
Since all important data structures contain type and size information, it is possible to:
\begin{itemize}
    \item mix several messages in a single stream,
    \item read new data with old code and vice versa, because messages with unknown tags can be skipped
    \item go through a file quickly until an event with a certain id is found, without reading and decoding all the other events into memory.
\end{itemize}
Only two different data structures are currently needed for handling all the different data of the experiment:

Shower data is stored in the \texttt{Event} structure, which has an id, a timestamp and an array of telescope events.
A telescope event has the telescope id, the pixel intensities, the times of maximum and the list of ids of significant pixels.
Events are stored in the Regions-Of-Interest file format (ROI)~\cite{ROI}, which allows to store the full camera image, a pixel list or a region of interest in the image.

Event parameterisations and subsystem data are stored in a \texttt{Parset} structure, which has an id, a timestamp and $n$ parameters, each of them being a vector.
Parameter sets are stored in FITS tables with 32 bit floating point precision.

\section{Disk storage}
\begin{figure}[h]
    \begin{center}
        \includegraphics[width=\textwidth]{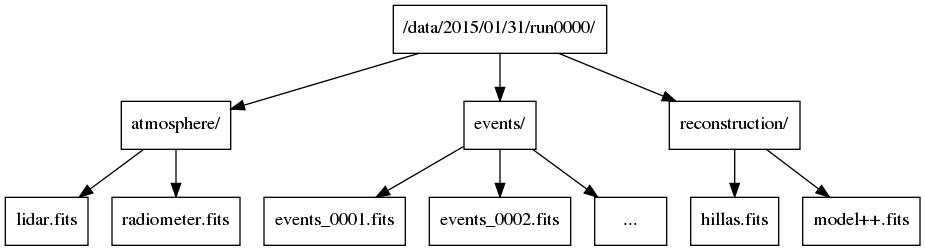}
    \end{center}
    \caption{Example of a directory structure suitable for MESS; events are stored in chunks.}
    \label{fig:dirtree}
\end{figure}

MESS requires the event data to be separated from event parameterisations and subsystem data.
This keeps the data structures, interfaces and file formats significantly simpler and allows users to access all that data with much less effort.
Different calibrations, results of updated reconstruction algorithms etc. are stored as additional extensions (HDUs) in the same FITS file.
This keeps the directories clean.
To prevent excessive file access, the results of the most frequent queries (like nightly, monthly and yearly summaries) can be stored in the respective directories.
Event data should be divided into chunks of length $\sim 1 s$, because that allows parallelisation by simply sending chunks to different computing nodes.
All files are stored in a directory structure similar to the one shown in figure \ref{fig:dirtree}.
By using this storage scheme instead of a database, the full power of the shell is at hand and it becomes easy to access the data.
The Hillas parameters of one night, for example, can be accessed with \texttt{/data/2015/01/31/*/reconstruction/hillas.fits}.

\section{On-the-fly selection}
Since parameterisations and subsystems are stored in FITS tables, the powerful column and row selection mechanisms of \textsc{cfitsio} can be used if the user wants to read only a subset of the data.
This way, most of the common queries can be done on the command line - without a database and without writing dedicated programs.
For example, if the user program shall read $\log(E)$ of all 3-telescope events, it is enough to write:

\begin{scriptsize}
    \begin{lstlisting}
    program -in "mc.fits[1][NVALID(d) == 3][col logE=log(mc_energy); d=mc_impact_dist]"
    \end{lstlisting}
\end{scriptsize}
Accessing vector columns is also possible.
This example shows how to get the impact distance of the fifth telescope for all events with an energy above 1 TeV:

\begin{scriptsize}
    \begin{lstlisting}
    program -in "mc.fits[1][mc_energy > 1][col mc_energy; d=mc_impact_dist[5]]"
    \end{lstlisting}
\end{scriptsize}
In both examples, the input file is filtered and transformed by \textsc{cfitsio} according to the expressions in the square brackets, and the user program then reads from that filtered table.

\section{Modules}
A module is the smallest functional unit in a MESS pipeline and it can have multiple inputs, process them and return multiple outputs.
It is defined in its own source code file and must expose at least an \textit{init}, an \textit{exec} and an \textit{exit} function.
From this, the shared object file is generated, which can then be dynamically loaded by the pipeline program.
When loaded, parameters can be passed to the module's \textit{init} function and there be accessed as \texttt{int argc, char **argv}, just like in a standalone program.
If there are more functions in the source file and if they are public (non-static), they are made available as library functions.
The module \texttt{hillas}, for example, has the three obligatory module functions \texttt{hillas\_init(...)}, \texttt{hillas\_exec(...)} and \texttt{hillas\_exec(...)},
but it also has the function \texttt{hillas\_televent(...)}, so the users can either define their pipeline with modules on the command line or write programs the traditional way: an executable calling libraries.

\section{Pipelines}
\begin{figure}
\begin{center}
    \includegraphics[width=\textwidth]{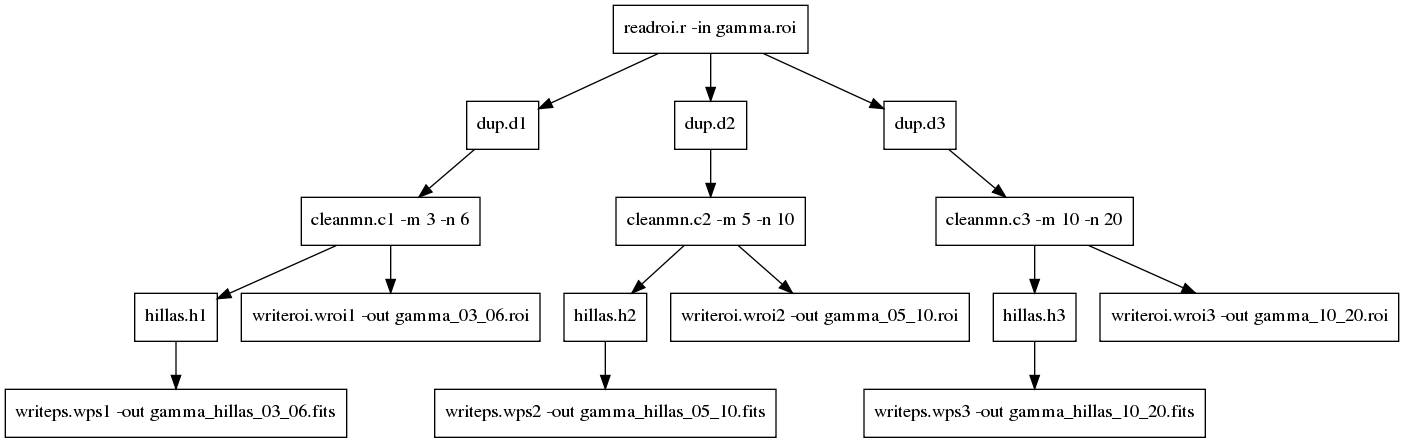}
\end{center}
\caption{Graph corresponding to the pipeline defined above.}
\label{fig:pipeline_hillas_graph}
\end{figure}

A MESS pipeline is a set of modules, which are executed in a defined order.
Each module can access the output of one or more other modules, but circular dependencies must be avoided.
Modules without parents usually read from files and then pass their data on to their children.
Modules without children usually write to files or display a plot.
Pipelines can be created on the command line by giving the types and names of the modules, their parent/child relations and their parameters.
The syntax for that is: \texttt{type.name:p1,p2,.. -par1 val1 -par2 val2 ...}, where p1, p2, ... is the list of parent modules to receive data from and 
\texttt{par1, val1, ...} are the parameter/value pairs of a module.
A depth-first topological sorting algorithm then resolves the dependency graph and returns the order in which the modules have to be initialised and executed.
Although nothing has to be compiled, it still runs as fast as a hand-written program containing the module calls in the correct order, because only pointers are passed between modules.
Even very complex pipelines covering the complete analysis chain can be easily defined on the command line or in scripts, without involving external libraries or threads.
Figure~\ref{fig:pipeline_hillas} shows an example of a MESS Pipeline that reads full-camera images, applies different cleanings, calculates the Hillas parameters and writes the results to disk,
with the corresponding graph shown in figure~\ref{fig:pipeline_hillas_graph}.

\begin{figure}
\begin{scriptsize}
    \begin{lstlisting}
mess -graph g1.dot -pipeline \
    readroi.r: -in gamma.roi , \
    dup.d1:r , dup.d2:r , dup.d3:r , \                                                                                                  
    cleanmn.c1:d1 -m 3 -n 6 , cleanmn.c2:d2 -m 5 -n 10 , cleanmn.c3:d3 -m 10 -n 20 , \
    hillas.h1:c1 , hillas.h2:c2 , hillas.h3:c3 , \
    writeps.wps1:h1 -out gamma_hillas_03_06.fits , \
    writeps.wps2:h2 -out gamma_hillas_05_10.fits , \
    writeps.wps3:h3 -out gamma_hillas_10_20.fits , \
    writeroi.wroi1:c1 -out gamma_03_06.roi , \
    writeroi.wroi2:c2 -out gamma_05_10.roi , \
    writeroi.wroi3:c3 -out gamma_10_20.roi ,
    \end{lstlisting}
\end{scriptsize}
\caption{Example of a MESS pipeline doing three different image cleanings, calculating the Hillas parameters and storing them to different files.}
\label{fig:pipeline_hillas}
\end{figure}

\section{Synchronisation}
\begin{figure}
    \begin{center}
    \includegraphics[width=\textwidth]{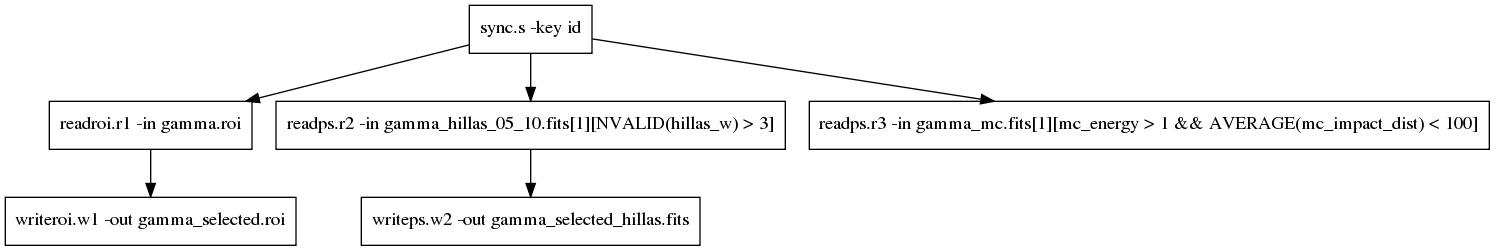}
\end{center}
\caption{Graph corresponding to the pipeline defined above.}
\label{fig:pipeline_sync_graph}
\end{figure}

Since each event and parameter set carries the global event time, it is possible to synchronise among different readers, for example for subsystem data, events or event parameterisations.
The \texttt{sync} module in MESS does this and it can be combined with the on-the-fly selection.
In the following example (see figures~\ref{fig:pipeline_sync} and \ref{fig:pipeline_sync_graph}), MC events, their Hillas parameters and their MC information are read from three different files and
synchronised such that only those events enter the pipeline that have 
Hillas parameters for more than three telescopes (\texttt{NVALID(hillas\_w) > 3}),
an energy of more than 1 TeV (\texttt{mc\_energy > 1}) and
a mean impact distance of less than 100 m (\texttt{AVERAGE(mc\_impact\_dist) < 100}):

\begin{figure}
\begin{scriptsize}
    \begin{lstlisting}
mess -graph g2.dot \
    -pipeline \
    sync.s: -key id , \
    readroi.r1:s -in gamma.roi , \
    readps.r2:s -in "gamma_hillas_05_10.fits[1][NVALID(hillas_w) > 3]" , \
    readps.r3:s -in "gamma_mc.fits[1][mc_energy > 1 && AVERAGE(mc_impact_dist) < 100]" , \
    writeroi.w1:r1 -out gamma_selected.roi , \
    writeps.w2:r2 -out gamma_selected_hillas.fits ,
    \end{lstlisting}
\end{scriptsize}
\caption{Example for a MESS pipeline using the synchronisation module.}
\label{fig:pipeline_sync}
\end{figure}

\newpage
\section{Plotting}
\label{sec:plotting}
\begin{figure}[ht!]
    \centering
    \includegraphics[width=0.49\textwidth]{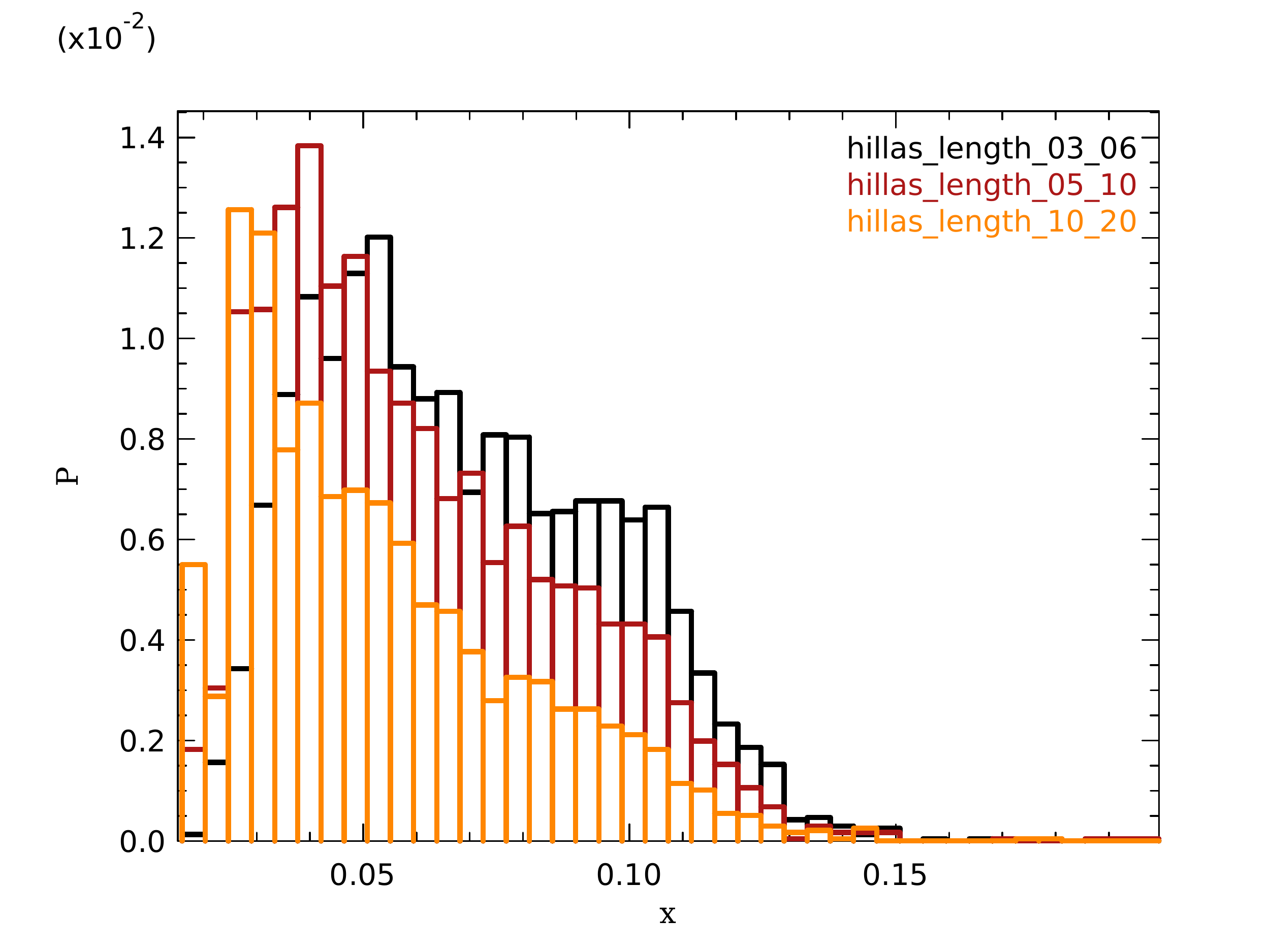}
    \includegraphics[width=0.49\textwidth]{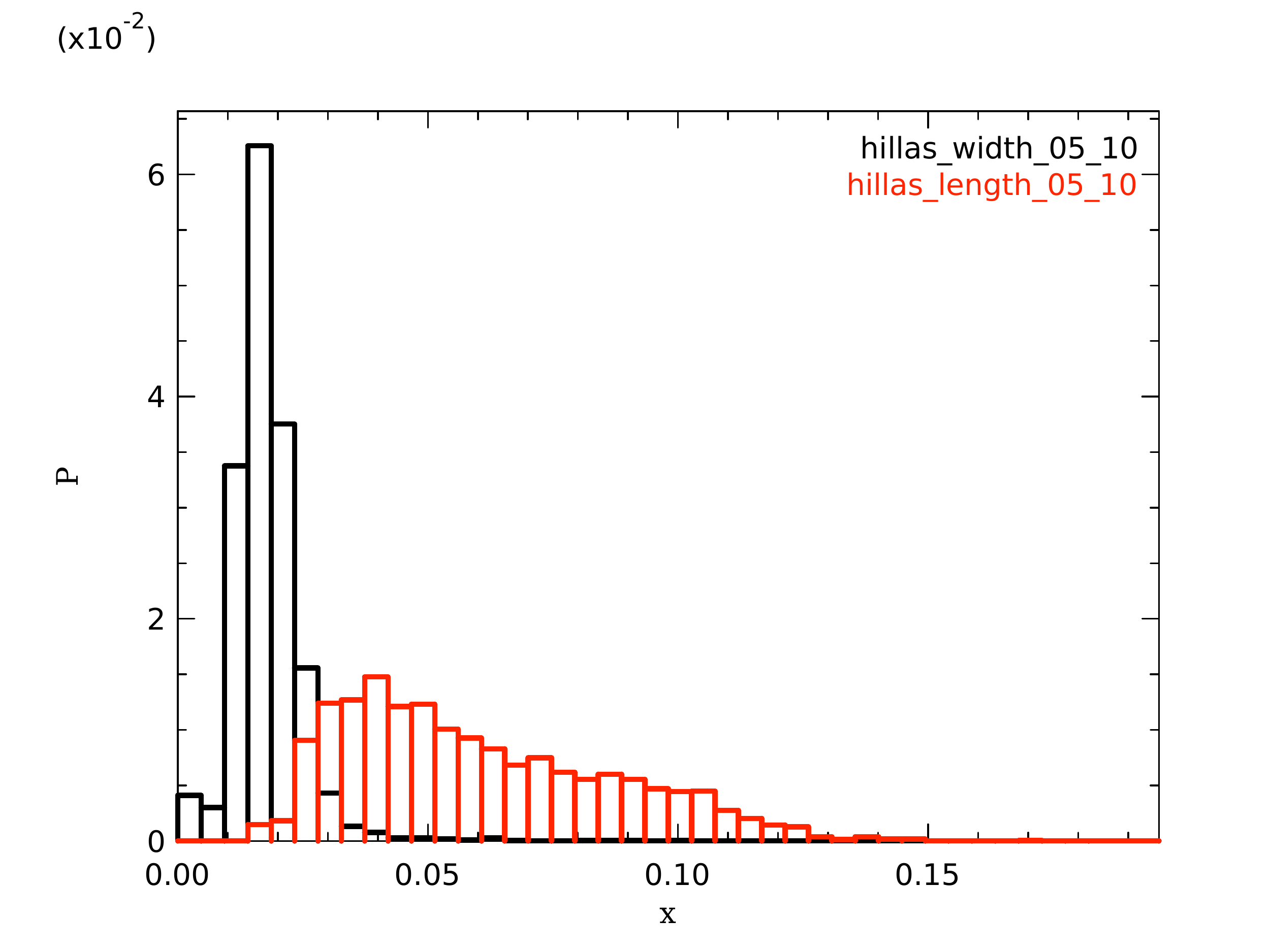}
    \caption{Hillas length for different image cleanings and Hillas width compared to Hillas length.}
    \label{fig:histograms}
\end{figure}

MESS provides a program to plot histograms of table columns of the FITS files given on the command line.
Through \textsc{cfitsio} column and row selection, the desired parameter and its range to be plotted can be specified.
The following two examples assume that the Hillas parameters of $3,6$-, $5,10$- and $10,20$-cleaned images have already been calculated.
Figure~\ref{fig:histograms} shows the three distributions of Hillas length on the left.
On the right, Hillas width and Hillas length of $5,10$-cleaned images are plotted.

\noindent\begin{minipage}[t]{0.49\textwidth}
    \begin{scriptsize}
        \begin{lstlisting}
mess.plothist -nbins 40 -in \
"gamma_hillas_03_06.fits[1][col hillas_l]"\
"gamma_hillas_05_10.fits[1][col hillas_l]"\
"gamma_hillas_10_20.fits[1][col hillas_l]"
        \end{lstlisting}
    \end{scriptsize}
\end{minipage}
\begin{minipage}[t]{0.49\textwidth}
    \begin{scriptsize}
        \begin{lstlisting}
  mess.plothist -in \
  "gamma_hillas_05_10.fits[1][col hillas_w]"\
  "gamma_hillas_05_10.fits[1][col hillas_l]"
        \end{lstlisting}
    \end{scriptsize}
\end{minipage}

\noindent

\section{Event display}
MESS provides the infrastructure for writing flexible event displays and supplies the user with a demo program using that functionality.
Displaying the whole array with the triggered telescopes and camera images is possible, as well as a detailed magnified view of the individual telescope events,
    showing either full camera images or only the regions of interest.
Each drawing routine can receive arbitrary parameters to draw event parameterisations on top of events, change the color palette etc.
The resulting images can then be exported to different formats, for example \textit{png}, \textit{eps} or \textit{pdf}.

\section{Conclusion}
MESS is a software framework designed for data processing in $\gamma$-ray astronomy, with emphasis on modularity, efficiency and simplicity.
It complies with the Unix philosophy and its programs can be easily embedded in scipts.
Its library allows developers to write modules and programs quickly, and with few lines of code.
The library functions can be used in C and C++ or wrapped for scripting languages like Python.

End users can define pipelines on the command line, which gives them much more flexibility than with config files, but without the need for programming or even scripting.
Several examples have shown how MESS pipelines can handle complex tasks that usually require writing a dedicated program.
Despite this flexibility, there is no degradation in performance or robustness, because MESS modules are shared libraries that are selectively pulled in.

Currently, MESS can read raw CTA MC data and perform all necessary steps to produce Gamma/Hadron separation plots from it, so more modules need to be developed for a complete CTA pipeline.
MESS is free software and can be downloaded from http://www.mpi-hd.mpg.de/~rmarx/mess.

\end{document}